\documentclass{article}

\date{}

\usepackage[letterpaper,margin=1in]{geometry}
\usepackage[parfill]{parskip}
\usepackage{microtype}
\usepackage{graphicx}
\usepackage{ifthen}
\usepackage{booktabs}
\usepackage{hyperref}
\usepackage{bm}
\usepackage{tikz}
\usepackage{amsmath}
\usepackage{amssymb}
\usepackage{mathtools}
\usepackage{amsthm}
\usepackage{physics}
\usepackage{colortbl}
\usepackage{xcolor}
\usepackage{float}
\usepackage{comment}
\usepackage{makecell}
\usepackage{caption}
\theoremstyle{plain}
\newtheorem{theorem}{Theorem}[section]

\theoremstyle{definition}
\newtheorem{definition}[theorem]{Definition}

\theoremstyle{remark}

\definecolor{darkblue}{rgb}{0, 0, 0.7}
\hypersetup{colorlinks,linkcolor={darkblue},citecolor={darkblue},urlcolor={darkblue}}

\def\cA{\mathcal{A}}

\def\cD{\mathcal{D}}
\def\cE{\mathcal{E}}

\def\cM{\mathcal{M}}

\def\cQ{\mathcal{Q}}
\def\cR{\mathcal{R}}

\def\cW{\mathcal{W}}

\def\Watermark{\mathsf{Watermark}}
\def\Detect{\mathsf{Detect}}
\def\Attribute{\mathsf{Attribute}}
\def\Decode{\mathsf{Decode}}
\def\prompt{\pi}
\def\gk{\mathsf{gk}} 
\def\dtk{\mathsf{dtk}} 
\def\dck{\mathsf{dck}} 
\def\ak{\mathsf{ak}} 
\def\m{\mathsf{m}} 
\def\true{\text{true}}
\def\false{\text{false}}

\usepackage[capitalize,noabbrev]{cleveref}

\usepackage{enumitem}

\newcommand{\xuandong}[1]{\textcolor{brown}{Xuandong: #1}}
\newcommand{\sam}[1]{\textcolor{red}{Sam: #1}}

\newcommand{\milad}[1]{\textcolor{violet}{Milad: #1}}

\newcommand{\yw}[1]{\textcolor{magenta}{yuxiang: #1}}

\usepackage{algorithm}
\usepackage{algorithmic}
\usepackage{authblk} 

\begin{document}

\title{SoK: Watermarking for AI-Generated Content} 

\renewcommand\Authfont{\small} 
\author[1]{Xuandong Zhao}
\author[1]{Sam Gunn}
\author[2]{Miranda Christ}
\author[1]{Jaiden Fairoze}
\author[3]{Andres Fabrega}
\author[4]{Nicholas Carlini}
\author[1]{Sanjam Garg}
\author[5]{Sanghyun Hong}
\author[6]{Milad Nasr}
\author[7]{Florian Tramer}
\author[8]{Somesh Jha}
\author[9]{Lei Li}
\author[10]{Yu-Xiang Wang}
\author[1]{Dawn Song}

\affil[1]{University of California, Berkeley}
\affil[2]{Columbia University}
\affil[3]{Cornell University}
\affil[4]{Anthropic}
\affil[5]{Oregon State University}
\affil[6]{Google DeepMind}
\affil[7]{ETH Zurich}
\affil[8]{University of Wisconsin–Madison}
\affil[9]{Carnegie Mellon University}
\affil[10]{University of California, San Diego}

\maketitle
\begin{abstract}
As the outputs of generative AI (GenAI) techniques improve in quality, it becomes increasingly challenging to distinguish them from human-created content. Watermarking schemes are a promising approach to address the problem of distinguishing between AI and human-generated content. These schemes embed hidden signals within AI-generated content to enable reliable detection. While watermarking is not a silver bullet for addressing all risks associated with GenAI, it can play a crucial role in enhancing AI safety and trustworthiness by combating misinformation and deception.
This paper presents a comprehensive overview of watermarking techniques for GenAI, beginning with the need for watermarking from historical and regulatory perspectives. We formalize the definitions and desired properties of watermarking schemes and examine the key objectives and threat models for existing approaches. Practical evaluation strategies are also explored, providing insights into the development of robust watermarking techniques capable of resisting various attacks. Additionally, we review recent representative works, highlight open challenges, and discuss potential directions for this emerging field. By offering a thorough understanding of watermarking in GenAI, this work aims to guide researchers in advancing watermarking methods and applications, and support policymakers in addressing the broader implications of GenAI.
\end{abstract}


\section{Introduction}
\label{sec:intro}

Generative AI (GenAI) techniques have become increasingly advanced~\cite{openai2022chatgpt, openai2023dalle3, openai2024sora, team2023gemini, ho2020denoising, rombach2022high}, transforming how content is created across a wide range of fields, from education~\cite{lewis2023generationai} and software development~\cite{github2021copilot} to creative industries~\cite{midjourney2022} and biological sciences~\cite{jumper2021highly}. Yet, the growing ability of these models to produce highly realistic and convincing content has also raised critical concerns about authenticity, attribution, and potential misuse~\cite{grinbaum2022ethical, crothers2023machine, yang2023survey, pei2024deepfake}. Distinguishing between human-created and AI-generated content is becoming increasingly challenging, necessitating the development of effective techniques to maintain transparency, accountability, and the responsible use of GenAI.

One possible solution to this challenge would be simply keeping a record of all content generated by an AI model. Such records could be retrieved to verify whether any piece of content is AI-generated~\cite{krishna2024paraphrasing}. This solution suffers from a number of limitations: (1) it requires substantial storage and cross-organizational coordination; (2) it raises privacy concerns as the record could expose private interactions with AI models; and (3) it cannot be applied to open-source models.

Another alternative is post-hoc detection of AI-generated content~\cite{korshunov2018deepfakes, mitchell2023detectgpt, bao2023fast, hu2023radar, hans2024spotting}, often relying on (possibly learned) statistical features to differentiate between human-created and AI-generated content. In the early stages of AI model development, these methods showed promise---generated images often exhibited identifiable artifacts, such as inconsistencies in rendering hands or compositional errors, while generated text contained distinct stylistic markers. However, as GenAI techniques have evolved, their outputs have become increasingly realistic, and these superficial clues have largely disappeared. The boundary between human and GenAI output is now blurred, and detection tools that rely on such inherent differences struggle to keep pace~\cite{liang2023gpt, tufts2024exam, shi2024red}. For instance, OpenAI's \emph{AI classifier}~\cite{openai2023classifier}, designed to detect AI-written text, was eventually deprecated due to its low accuracy and inconsistent performance, underscoring the limitations of post-hoc approaches.

\begin{figure}[t]
    \centering
    \includegraphics[width=1.0\linewidth]{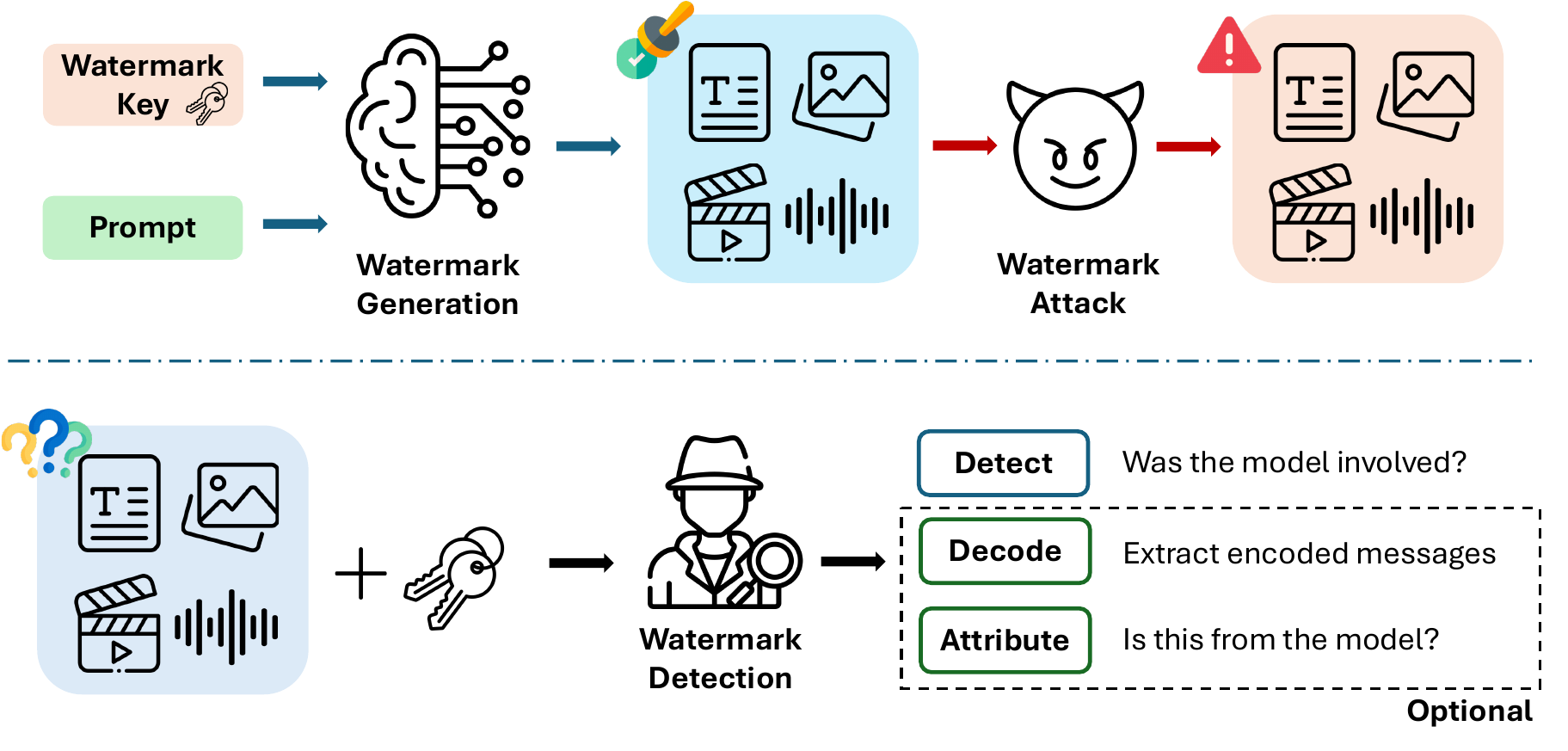}
    \caption{Outline of the watermarking scenario. This figure illustrates the core components of watermarking for GenAI, including watermark generation, detection, and potential attacks. See \Cref{sec:watermark_method} for a detailed explanation of the properties desired of a watermarking scheme.}
    \label{fig:main}
\end{figure}

\emph{Watermarking}~\cite{aaronson, kirchenbauer2023watermark, zhao2024provable, christ2024undetectable, wen2023tree, gunn2024undetectable, chen2023wavmark, san2024proactive} offers a potentially more robust and reliable solution to this problem. Unlike methods that passively attempt to detect incidental differences between human and AI-generated content, watermarking involves actively embedding an imperceptible signal within GenAI output at the time of its generation. This watermarking process subtly modifies the output while maintaining its intended meaning and appearance, embedding a hidden but consistent signal that serves as a verifiable marker of GenAI origin (see Figure \ref{fig:main} for an overview). This technique is not constrained by the evolving capabilities of GenAI models but instead provides a deliberate and persistent means of identification. The watermark can be detected later, allowing for verification without relying on inherent or statistical differences, thereby offering a more resilient way to distinguish machine-generated content from human work~\cite{yang2023survey, an2024waves, jiang2024watermark}\footnote{Watermarking can also facilitate the decoding of embedded long messages, enabling attribution to specific models (discussed in Section \ref{sec:watermark_method}). Moreover, watermarking principles extend to other domains, such as model and dataset watermarking, which are briefly addressed in Section \ref{sec:open_prob}. However, the primary focus of this paper is the detection of GenAI content.}.

This paper presents a comprehensive overview of watermarking techniques for AI-generated content. 
The organization of the paper reflects our key contributions, structured as follows: Section~\ref{sec:why_watermark} examines the historical context and regulatory landscape driving the need for watermarking in the GenAI era. Section \ref{sec:watermark_method} introduces the terminology, notation, and syntax of watermarking schemes, followed by essential properties and definitions that characterize effective watermarks. Specifically, a successful watermarking scheme should maintain model output quality, achieve high detection accuracy, exhibit robustness against evasion attacks, resist forgery, and remain computationally efficient. Section \ref{sec:threat_model} analyzes various threat models and attack techniques that watermarking schemes might encounter, providing a detailed analysis of security challenges. Section \ref{sec:eval_method} outlines practical methods for evaluating the effectiveness and resilience of watermarking schemes in diverse applications. Section \ref{sec:represent_works} surveys recent representative methods across different modalities to illustrate the evolution of existing approaches. Finally, Section \ref{sec:open_prob} explores open problems in watermarking GenAI content, discusses current limitations, and identifies promising future directions.

Overall, while watermarking is not a comprehensive solution to all challenges associated with GenAI, it represents a critical tool for enhancing AI safety and fostering trust. Through this paper, we aim to provide practitioners, researchers, and policymakers with a thorough understanding of the current state of watermarking in generative AI, while highlighting key areas for future investigation. 

\subsection{Methodology for Systematizing Knowledge}
To construct this SoK, we conducted an extensive review of over 100 peer-reviewed papers published from the early 2000s to the 2020s. Our sources include top venues in security (e.g., USENIX Security, IEEE S\&P), AI (e.g., NeurIPS, ICML), and multimedia (e.g., CVPR, ACM Multimedia), as well as relevant arXiv preprints reflecting the fast-paced development of the field. We further incorporated key regulatory and industry documents, such as the EU AI Act and C2PA specifications. Our selection criteria prioritized works by their relevance to GenAI watermarking across various modalities, contribution to definitions, threat modeling, evaluation, or technical advances, and identification of unresolved issues and future directions. This structured approach enabled impartial and accurate representation of diverse perspectives and established principles within the surveyed literature.


\subsection{Comparison with Prior Surveys}
While watermarking has a long history, its application to GenAI content is a relatively recent and rapidly developing field. Several surveys have touched upon related aspects. Traditional digital watermarking surveys (e.g., \cite{cox2007digital}) provide foundational concepts but predate modern GenAI. Surveys on deep learning watermarking often focus on \emph{model watermarking} for ownership protection (e.g., \cite{li2021survey, boenisch2021systematic}), which has different goals and technical challenges than content watermarking. Amrit and Singh~\cite{amrit2022survey} provide a broad overview of watermarking in AI but with limited focus on generative models specifically. 

In contrast, our SoK offers a timely and in-depth analysis focused specifically on watermarking outputs of GenAI models. Our key contributions are: (1) Scope: Coverage across multiple modalities—text, image, audio, and video. (2) Depth: Formal definitions, structured threat models, and evaluation criteria. (3) Timeliness: Inclusion of recent advances, such as undetectable and provable watermarking schemes. Together, these elements position our paper as a comprehensive and authoritative resource on watermarking in the GenAI era.

\section{Why Watermark GenAI Content?} \label{sec:why_watermark}

In this section, we begin by highlighting the limitations of post-hoc detection methods for identifying AI-generated content. Next, we provide a brief history of watermarking and discuss its heightened importance in the era of GenAI. We then outline the key use cases and motivations for implementing GenAI watermarks. Finally, we examine current policies related to GenAI watermarking in both government and industry. Together, these points underscore the critical need for watermarking AI-generated content.
\subsection{Limitations of Post-hoc Detection}
Post-hoc detection methods for identifying AI-generated content can be broadly classified into two categories: zero-shot detection~\cite{mitchell2023detectgpt, bao2023fast, hans2024spotting} and training-based detection~\cite{korshunov2018deepfakes, hu2023radar}. In zero-shot detection, the detector leverages statistical signatures that are characteristic of AI-generated content. On the other hand, training-based detection involves training a binary classification model on datasets that include both human and AI-generated content.

Watermarking techniques actively add signals to ensure reliable detection, while post-hoc detection methods adopt a passive approach. Despite their versatility, post-hoc detection methods often exhibit low performance. The reported error rates are only validated empirically on limited datasets and typically do not fall below $10^{-3}$~\cite{hans2024spotting, giboulot2024watermax}. Additionally, post-hoc detection methods cannot provide theoretical guarantees for the false positive rates and face fundamental limitations when handling out-of-distribution data~\cite{liang2023gpt}.
 
As generative models continue to advance, their outputs are increasingly realistic, sometimes even surpassing the quality of human-generated content. This makes previously trained classifiers less effective for detecting outputs from newer generative models. Furthermore, most existing post-hoc detection approaches aim to distinguish between human and AI-generated content without providing more granular information, such as model versions or user-specific details. These limitations make post-hoc detection suboptimal for detecting AI-generated content effectively.

\subsection{History of Watermarking} 
Watermarking has a long historical background, rooted in the need to authenticate, protect, and trace various types of content. Its origins can be traced back to the early 13th century~\cite{hunter1916hand}, when it was first employed in paper production. Thin wire patterns were added to paper molds, creating physical marks that identified the manufacturing mill or indicated the paper's quality, while sometimes serving as decorative elements~\cite{cox2007digital}. By the 18th century, watermarking had become an essential tool in anti-counterfeiting efforts, particularly in currency production, where it was used to deter forgery and verify authenticity.

With the advent of digital technology, watermarking evolved to protect intellectual property, trace sources, and authenticate media in the digital domain~\cite{van1994digital}. Musicians, filmmakers, and artists began using digital watermarks to secure their work from unauthorized copying and redistribution. In digital media, watermarks could be embedded into images, audio, and video files to establish ownership or copyright, either as visible markers or as hidden, verifiable information embedded through algorithms. This approach helped combat piracy while upholding the rights of content creators. 

The historical importance of watermarking lies in its dual role as a preventive and protective tool. From printed currency to copyrighted media, watermarking has consistently preserved authenticity and safeguarded value. This historical evolution provides essential context for understanding watermarking's role in the current GenAI era, where it serves similar purposes in new the digital world. 

In the GenAI era, debates surrounding the copyright status of AI-generated data have become increasingly prominent. The focus of this paper is on watermarking the output of GenAI, specifically in the context of detecting AI-generated content. Whether such outputs should be considered copyrighted material remains an ongoing discussion. Besides, we mainly focus on invisible watermarks, which are embedded in such a way that they are imperceptible to human eyes but can be detected by a specialized algorithm. Visible watermarks are often superimposed onto images or video content. They include logos, text, or symbols, which clearly identify the creator or ownership of the content. While effective for discouraging unauthorized use, visible watermarks can be aesthetically intrusive and may be removed by skilled users or automatic tools~\cite{huang2004attacking, dekel2017effectiveness}.

\subsection{The Needs of Watermarking in the GenAI Era} 
Watermarking has historically played a significant role and has become even more critical in addressing the unique challenges of GenAI. Below, we outline key use cases and motivations for implementing watermarking systems in the GenAI era.

\noindent\textbf{Combating Misinformation.}
GenAI can be exploited to generate misinformation at a large scale.
Adding watermarks to AI-generated content could help trace the source of the information and, therefore, facilitate identifying misinformation and disinformation. 
At the least, it would help readers distinguish human-created content from AI-generated material if all AI content carried an AI flag.

\noindent\textbf{Enhancing Fraud Detection.}
While current scam campaigns are largely automated, they typically involve replicating similar messages to numerous recipients. This generic approach makes such content easily detectable. 
However, generative models could produce tailored scam messages, making traditional detection methods less effective. Watermarking could play a crucial role in identifying and mitigating such sophisticated scam attempts, bolstering anti-scam efforts.

\noindent\textbf{Deterring Academic Dishonesty.}
There are already many cases in which students use ChatGPT~\cite{openai2022chatgpt} to complete homework assignments. 
Watermarking can help prevent cheating in domains where it's important to validate that text is human-written, such as student essays. By making it easier to identify AI-generated content, educational institutions can better ensure academic integrity.



\noindent\textbf{Avoiding Training Data Contamination.}
Machine learning models often depend on data scraped from the public internet for training. As AI-generated content becomes more prevalent online, detecting AI-generated data becomes vital. If models are inadvertently trained on their own generated outputs, it can lead to ``model collapse,'' where iterative cycles of training on degraded data degrade the overall quality of generated content~\cite{shumailov2024ai}. Watermarking can help to flag AI-generated content, allowing future models to prioritize high-quality human-created data, thereby preserving the integrity of training datasets.

\noindent\textbf{Providing Signatures and Attribution.}
Watermarks can serve as signatures, allowing users to verify that content was generated by a specific model. This is particularly useful for those concerned with using models trained on properly licensed data. Additionally, watermarks can help prove whether harmful content attributed to a model was genuinely generated by that model or not.

By addressing these needs, watermarking emerges as a critical tool for fostering responsible AI development, building trust, and mitigating misuse. While this paper primarily focuses on detecting GenAI content, other watermarking applications, such as \emph{model watermark} and \emph{dataset watermark}, are briefly discussed in Section~\ref{sec:open_prob}.



\subsection{Policies in Government and Industry} 

Watermarking AI-generated content has garnered significant attention from both governments and industries, highlighting its importance for responsible AI governance. By instituting policies that mandate watermarking, both government bodies and private companies are taking important steps toward minimizing the risks associated with GenAI content, while also empowering end users with the tools needed to discern the origins and authenticity of the content they encounter. This section summarizes the major developments.

\subsubsection{Overview of Government Developments}

Governments around the world are increasingly recognizing the importance of watermarking as a regulatory tool to combat misinformation and promote transparency, resulting in various policy developments. We summarize a few notable examples below.

\noindent\textbf{United States.} 
In October 2023, the White House issued Executive Order 14110~\cite{eo14110}, which introduces a set of guidelines for the executive branch related to GenAI content. A component of this order mandates that a number of government agencies produce a report within 240 days identifying state-of-the-art techniques for detecting, labeling (through methods such as watermarking), and tracking synthetic~(e.g., GenAI) content. This report will be used to develop guidance for government agencies regarding detection of GenAI content. It was also noted that the Department of Commerce will take the lead in this effort.

In addition to this executive order, there have been other legislative efforts related to GenAI watermarking. At the state level, a notable example is California, which introduced the California AI Transparency Act~(SB 942)~\cite{sb942}. This bill---which was signed into law by the governor, and is set to take effect in 2026---mandates that providers of GenAI models must release publicly-available tools to detect whether image, video, or audio content was created or modified by their models. Additionally, providers must include a ``latent disclosure'' in images, videos, and audio generated by their models, which must be compatible with their detection tools.

In addition to SB 942, California also proposed bill AB 3211~\cite{ab3211}, which included additional provenance-tracking requirements alongside other GenAI-related provisions, though it was ultimately rejected by the State Senate. Other states, such as Ohio~(senate bill 217~\cite{ohio217}), have also taken preliminary steps to introduce laws that would mandate watermarks on GenAI content.

At the federal level, several bills have been proposed by members of Congress, including the Advisory for AI-Generated Content Act (S.2765)~\cite{s2765}, the AI Labeling Act~(S.2691)~\cite{s2691}, the COPIED Act~\cite{copiedact}, and the Artificial Intelligence Research, Innovation, and Accountability Act~(S.3312)~\cite{s3312}. Among other things, mandates across these bills include requirements for inserting watermarks and disclosures on outputs of GenAI models, and requests for the development of standards and recommendations for detection of GenAI content. 

\noindent\textbf{European Union.} In the EU, the preeminent policy development related to watermarking is the EU AI Act~\cite{eu-ai-act}, which came into effect in August 2024. This comprehensive regulatory framework addresses a myriad of topics aimed at mitigating the risks associated with GenAI content. Most relevant to watermarking, Article 50 includes transparency provisions requiring that GenAI providers ensure that outputs of their models are ``marked in a machine-readable format'', enabling the detection of digital content generated or manipulated by their models. Recital 133 provides additional context for these provisions, explicitly mentioning watermarks and ``cryptographic methods'' as potential techniques for marking such content.

The specific, practical implementation of the requirements outlined in Article 50 will be based on forthcoming guidelines from the European Commission, as specified in Article 96.

\noindent\textbf{Other Government Regulations.} Several other countries have also introduced regulatory measures to address challenges posed by AI-generated content. In China, the Provisions on the Administration of Deep Synthesis Internet Information Services~\cite{china-ai-provisions} mandates that model providers include notable labels on GenAI content that may ``cause confusion or mislead the public'' (Chapter 3, Article 17). Similarly, South Korea's Content Industry Promotion Act~\cite{sk-ai-provisions} seeks to support the content industry while safeguarding creators' rights. 
To address issues arising from AI-generated content, South Korea has proposed a revised bill aimed at establishing clear legal frameworks for distinguishing and regulating such outputs, including text, images, and music.

\noindent\textbf{Summary.} These policy developments around the world evidence the fact that regulatory bodies have a strong interest in identifying GenAI content, which they see as an important component of AI safety. However, the vagueness of these legal documents underscores the importance of a deeper understanding of the technical capabilities of watermarking schemes. As more jurisdictions develop their regulatory frameworks for GenAI watermarking, it is important for laws to ``align with the capabilities and limitations of current watermarking technology''~\cite{christodorescu2024securing}, and be cognizant of the differences between existing watermarking techniques. 

\subsubsection{Overview of Industry Developments}
In the industry sphere, major technology companies such as Google~\cite{synthid2024, dathathri2024scalable}, OpenAI~\cite{aaronson, Bartz_Hu_2023}, and Microsoft~\cite{ms2023} have started to implement watermarking systems voluntarily, either as standalone initiatives or as part of broader AI governance frameworks. These companies recognize that watermarking is an important component of responsible AI use, enabling better content moderation and ensuring that their generative models are not exploited to produce harmful or misleading content. Additionally, industry consortiums, like the Partnership on AI, have called for standardized best practices around the watermarking of AI content, ensuring cross-platform coherence and effectiveness. We discuss a few important developments below.

\noindent\textbf{C2PA.} The Coalition for Content Provenance and Authenticity (C2PA)~\cite{c2pa}, a collaboration between the Content Authenticity Initiative (CAI) and Project Origin, is a group of companies and other stakeholders that seeks to establish standards for digital media provenance. CAI, founded by Adobe, the New York Times, and Twitter, focuses on enabling creators to establish authorship while empowering consumers to assess content reliability. Project Origin, founded by BBC, CBC Radio Canada, Microsoft, and the New York Times, aims to combat disinformation by developing methods to affirm content integrity.

Content provenance offers a method to verify the origin of digital content, including information on its creator, creation time, and edits~\cite{c2pa-spec}. Unlike metadata, which can be manipulated, provenance data provides a reliable record, allowing users to distinguish authentic content from manipulated or synthetic material. By embedding trustworthy origin data, provenance reduces misinformation risks, enhancing transparency in digital content distribution at scale.

\noindent\textbf{Google DeepMind SynthID.} SynthID~\cite{synthid2024, dathathri2024scalable} is a pioneering technology developed by Google to address the challenge of identifying AI-generated content across various media formats, including text, audio, images, and video. By embedding imperceptible digital watermarks directly into AI-generated outputs, SynthID ensures content verification without compromising quality. The technology uses advanced deep learning models to adjust token probabilities in text, modify spectrograms in audio, and embed watermarks into image pixels or video frames, making the markers resilient to common alterations such as compression, cropping, or noise addition. SynthID’s applications are broad, ranging from text generation via the Gemini platform~\cite{gemini-text-api} to audio production with Lyria~\cite{lyria-api}, as well as image and video generation AI models like Imagen~\cite{ho2022imagen} and VideoFX~\cite{videofx}. Google recently released a public implementation of their text watermarking scheme~\cite{synthid-text-watermark}, which will facilitate applying these techniques to other models.

\section{What is a Watermark?} \label{sec:watermark_method}
There are many different properties one might desire of a watermarking scheme, the relative importance of which depends on the application scenario.
In this section, we will outline what we view as the most important of these properties.

We begin by introducing terminology, notation, and the syntax of a watermarking scheme in general.
We have attempted to keep the presentation as simple as possible, without significantly compromising on generality or precision.

Let $\cM$ be any generative model.
To denote the process of sampling a response $x$ from the model on input prompt $\prompt$, we write $\cM(\prompt) \to x$. 
For instance, $x$ could be an image, a sequence of text, an audio sample, or a video.

The central piece of any watermarking scheme is the watermark generation algorithm $\Watermark_{\gk}^{\cM}$.
This algorithm takes as input a prompt, and uses the generative model $\cM$ and the watermark generation key $\gk$ to produce a watermarked response $x$.
We denote this process by $\Watermark_{\gk}^{\cM}(\prompt) \to x$.
A ``standard'' watermark that simply embeds a detectable signal and no further message is called a \emph{zero-bit watermark}.
For \emph{multi-bit watermarks}, $\Watermark_{\gk}^{\cM}$ also takes as input a message $\m \in \{0,1\}^k$ to be embedded in the generation, and we write $\Watermark_{\gk}^{\cM}(\m, \prompt) \to x$.

In order for the watermark to be useful, the watermarking scheme must additionally come with one or more of the following algorithms:
\begin{itemize}
    \item $\Detect_{\dtk}(x) \to \{ \true, \false \}$ tests content $x$ for the watermark. If $\Detect_{\dtk}(x) = \true$, then $x$ is not independent of $\dtk$.
    \item $\Decode_{\dck}(x) \to \{0,1\}^k$ recovers the message from the content $x$, if there is one.
    \item $\Attribute_{\ak}(x) \to \{ \text{true, false} \}$ tests content $x$ for an \emph{exact} match with a watermarked generation. While $\Detect$ is used, for instance, to identify misinformation being spread by malicious \emph{users} of the model, $\Attribute$ is used to accuse the model \emph{owner} of producing problematic content. See \Cref{subsec:unforgeability} for a more in-depth discussion of the differences between detection and attribution, and why they cannot be the same function.
\end{itemize}
We have allowed for separate generation ($\gk$), detection~($\dtk$), decoding ($\dck$), and attribution ($\ak$) keys, but in many schemes some or all of these are identical.

\subsection{Quality}
Ideally, the watermark should not degrade the quality of the model in any way.
Watermarking works define and achieve various relaxations of this goal, and we present these quality notions from weakest to strongest.

These notions differ in two especially notable ways: heuristic versus provable, and single-response versus multi-response.
Heuristic guarantees, namely empirical quality validation, are limited by the scope of the experiments they involve; for example, the watermark may significantly harm quality on prompts or tasks not captured in the experiments. 
Furthermore, the quality metric used in experiments may be biased or unreliable.
On the other hand, the provable quality notions we present apply to \emph{any possible prompt}, and ensure that any bias introduced by the watermark is limited, or even nonexistent.

Second, some guarantees (e.g., some empirical quality validation methods, low-distortion, and distortion-freeness)
 apply only to a \emph{single} response from the watermarked model. 
On the other hand, other notions (e.g., stronger empirical quality validation methods and undetectability) show that \emph{any number} of watermarked responses are \emph{jointly high-quality}.
This distinction is important in ensuring the watermark does not harm variability of the model, or introduce systematic biases.
For example, an image watermark may cause a model to output exclusively images of dogs, but which are extremely high-quality.
This watermark would satisfy the single-response quality notions, since this bias would only appear given multiple responses.
This is especially problematic if the bias is more insidious, or if the model is being used for complicated downstream tasks involving multiple outputs.

\subsubsection{Empirical Quality Validation}
These empirical measurements include prompting other models to score the quality of watermarked content, asking users to compare the quality of watermarked versus unwatermarked content, or measuring the performance of downstream tasks that use model-generated content.
Such experiments can involve many responses from the model and can therefore yield (empirical) multi-response quality guarantees.
However, these empirical measures are not representative of \emph{all} uses.
For example, the experiments may overlook a class of prompts on which the watermark yields poor quality outputs.

In particular, for production-scale models the quality tests run on the original model are extremely cost- and resource-intensive.
The empirical quality tests run in watermarking works do not approach this scale, with the exception of SynthID~\cite{synthid2024}\footnote{SynthID in particular use human evaluation to evaluate the effect of watermarking over 5 categories of grammar, relevance, correctness, helpfulness, and overall quality. They also empirically evaluated the detection rate of their proposed approach and showed it has low false positive rates.}.
As a result, watermarks with \emph{only} empirical quality guarantees are less likely to be fit for widespread use. 
However, empirical quality validation is often used in conjunction with the provable quality guarantees detailed below.

\subsubsection{Low-distortion and Distortion-free Watermarks}
We define the \emph{distortion} as the maximum, over all model inputs, of the statistical distance between the watermarked and unwatermarked response distributions.

\begin{definition}[Distortion]
Formally, the distortion is
\ifthenelse{\boolean{@twocolumn}}{
\begin{multline*}
    \max_{\m,\prompt} \frac{1}{2} \sum_{x \in \cR} \Big| \Pr[\cM(\prompt) \to x] \\- 
    \Pr_{\gk}[\Watermark_{\gk}^{\cM}(\m, \prompt) \to x] \Big|
\end{multline*}
}{
\[
\max_{\m,\prompt} \frac{1}{2} \sum_{x \in \cR} \Big| \Pr[\cM(\prompt) \to x] - 
    \Pr_{\gk}[\Watermark_{\gk}^{\cM}(\m, \prompt) \to x] \Big|
\]
}
where $\cR$ is the set of possible (partial) responses and $\m$ is a message encoded by the watermark.
\end{definition}

Some language model watermarks, such as that of Green-Red Watermarks~\cite{kirchenbauer2023watermark, zhao2024provable}, achieve \emph{low distortion} over individual token distributions.
Multiple or long watermarked responses may jointly exhibit greater degradation, and therefore low-distortion is fairly weak.
For other modalities, where content is output in its entirety rather than in discrete units such as tokens, one typically takes $\cR$ to be the set of entire responses.

Kuditipudi et al.~\cite{kuditipudi2023robust} showed that for language models, is possible to obtain build schemes that achieve zero distortion for the distribution of an entire single response; such schemes are called \emph{distortion-free}.
While originally defined for text, distortion-freeness is achievable across all modalities, including images~\cite{yang2024gaussian}.
Distortion-freeness can be defined either computationally or statistically, and we present both definitions.

\begin{definition}[Distortion-freeness~\cite{kuditipudi2023robust}]
Formally, a watermark is \emph{computationally distortion-free} if for any prompt $\prompt$, watermark message $\m$, security parameter $\lambda$, and polynomial-time algorithm $D$,\footnote{The notation $\mathsf{negl}(\lambda)$ denotes a function that decays faster than any inverse polynomial in the security parameter $\lambda$.
The watermarking keys should have length that is dependent on $\lambda$, but we leave this implicit here.
We give $1^\lambda$ to $D$ to ensure that $D$ is allowed to run in any time that is polynomial in $\lambda$.
This convention comes from cryptography, where $\lambda$ is chosen such that $\mathsf{negl}(\lambda)$ is so small a probability that in practice this event never occurs, hence the term \emph{negligible}.}
\ifthenelse{\boolean{@twocolumn}}{
    \begin{multline*}
        \Big| \Pr_{x \gets \cM(\prompt)}\left[D^{\cM}(1^{\lambda}, x) \to 1 \right] \\ 
        - \Pr_{\substack{\gk \\ x \gets \Watermark_{\gk}^{\cM}(\m, \prompt)}}\left[D^{\cM}(1^{\lambda}, x) \to 1 \right] \Big| \leq \mathsf{negl}(\lambda).
    \end{multline*}
}{
    \[
        \Big| \Pr_{x \gets \cM(\prompt)}\left[D^{\cM}(1^{\lambda}, x) \to 1 \right] 
        - \Pr_{\substack{\gk \\ x \gets \Watermark_{\gk}^{\cM}(\m, \prompt)}}\left[D^{\cM}(1^{\lambda}, x) \to 1 \right] \Big| \leq \mathsf{negl}(\lambda).
    \]
}
If the above holds even for computationally unbounded algorithms $D$, the watermark is \emph{statistically distortion-free}.
\end{definition}

In other words, the algorithm $D$ is given a single response from either the original model, or the watermarked model. 
It may make additional queries to the original model, then output either 0 or 1 to indicate whether it believes the given response was watermarked.
Distortion-freeness requires that in either case, the probability that $D$ outputs 1 is the same up to a negligible additive term.

Computational distortion-freeness is generally no weaker than statistical distortion-freeness in practice, as real-world algorithms are efficient.
Distortion-free schemes ensure that the quality is preserved for any \emph{single} response from the model, but do not guarantee anything about the quality across multiple generations.
By simply using multiple pairs of generation and detection/decoding keys, it is possible to make any distortion-free scheme satisfy a low-distortion notion for multiple responses.
However, this comes at the expense of the detector's computational efficiency and the false positive rate.

\subsubsection{Undetectable Watermarks}
A watermarking scheme is \emph{undetectable} if it is computationally infeasible to distinguish between the output distributions of the original model $\cM$ and the watermarked model without the keys, even when \emph{adaptive queries} are permitted.
In particular, unlike the extension of distortion-freeness, the number of queries need not be known in advance.
We define only the computational version of undetectability, as it is impossible to achieve statistical undetectability~\cite{christ2024undetectable}.

\begin{definition}[Undetectability~\cite{christ2024undetectable}]
A watermarking scheme is \emph{undetectable} if, for every security parameter $\lambda$ and every polynomial-time algorithm $D$:
\ifthenelse{\boolean{@twocolumn}}{
    \begin{multline*}
    \Big| \Pr\left[D^{\cM, \cM}(1^\lambda) \rightarrow 1\right] \\- 
    \underset{\gk}{\Pr} \left[D^{\cM, \Watermark_{\gk}^{\cM}}(1^\lambda) \rightarrow 1\right] \Big| \leq \mathsf{negl}(\lambda),
    \end{multline*}
}{
 \[
    \Big| \Pr\left[D^{\cM, \cM}(1^\lambda) \rightarrow 1\right] - 
    \underset{\gk}{\Pr} \left[D^{\cM, \Watermark_{\gk}^{\cM}}(1^\lambda) \rightarrow 1\right] \Big| \leq \mathsf{negl}(\lambda),
 \]
}
where the notation $D^{\mathcal{O}_1, \mathcal{O}_2}$ means that $D$ has adaptive query access to the oracles $\mathcal{O}_1$ and $\mathcal{O}_2$ using arbitrary prompts and watermark messages.
The oracle for the original model $\cM$ responds to the prompt but ignores the given watermark message.
\end{definition}

In other words, $D$ attempts to determine whether it has query access to only the original model, or to both the original model and the watermarked model. 
An alternate and formally equivalent definition allows $D$ to access only one of the original or watermarked model at a time.

Although it was originally defined for language models, undetectability applies generally to any modality and is achievable for image watermarks~\cite{gunn2024undetectable}.

If a scheme is undetectable, this implies that it preserves the quality of the original model under \emph{any efficiently computable quality metric}; otherwise, such a metric would serve as a distinguisher.
This holds even for quality metrics that measure quality across many generations.
Undetectability is therefore the most powerful quality guarantee, ensuring that the watermark does not harm performance on any possible downstream task.
It also implies that it is impossible to learn detection, decoding, or attribution keys, which can sometimes help prevent watermark removal attacks.


\subsection{False Positive Rate}
A crucial property of the detector for watermarking scheme is that it has a low false positive rate.
That is, it should be highly unlikely for our detector to flag \emph{any content} that is produced independently from the watermarking keys.

\begin{definition}[False positive rate]
    A watermark detector $\Detect$ has false positive rate at most $\varepsilon$ if, for any fixed content $x$,
    \[
        \Pr_{\dtk}[\Detect_{\dtk}(x) \to \text{true}] \le \varepsilon.
    \]
\end{definition}

Note that the above definition is agnostic to the distribution of natural content.
This is absolutely essential for the interpretability and credibility of the watermark detector.
If the false positive rate only holds for content generated under certain conditions, then it is possible that the detector will disproportionately flag some kinds of content.

\subsection{False Negative Rate and Robustness}
The false negative rate describes how reliably one can detect the watermark in \emph{unmodified} watermarked content.
The false negative rate depends not only on the watermarking scheme, but also on the amount of randomness in the content being generated.
For example, if the model is prompted to output a fixed response, one cannot hope to embed the watermark, unless one gives up \emph{distortion-free}ness (which means not faithfully following instructions in the ``fixed-response'' case).
Therefore, provable false negative guarantees often state that the watermark is detectable with high probability \emph{in sufficiently random content}.
Christ et al.~\cite{christ2024undetectable} showed that such a condition is necessary for undetectable language model watermarks.
For other modalities where nearly all content is highly random, such as images and audio, this randomness condition is assumed rather than specified as a condition.

Robustness refers to the watermark's ability to withstand watermark removal attacks.
In practice, content may be edited, compressed, or otherwise altered.
For example, text may be paraphrased, while images may undergo resizing or compression.
A robust watermark should be resilient to such modifications, ensuring that it remains detectable even if the content undergoes transformations.

Unfortunately, definitions of robustness are significantly more complicated than those of the other properties described here.
There are several reasons for this:
\begin{itemize}
    \item Robustness must be defined with respect to a channel or class of channels. A channel $\cE$ models the action of the environment or an adversary attempting to remove the watermark. Importantly, it is not possible to be robust to every channel. For instance, any scheme with a low false positive rate cannot be robust to any channel which is input-independent.
    \item Robustness may depend on the \emph{knowledge} provided to the channel. It is sometimes possible to use the watermarking keys or detector/decoder access to craft stronger watermark removal attacks~\cite{jiang2023evading, jovanovic2024watermark, pang2024no}. We denote the channel applied to content as $\cE(x)$, leaving the possible dependence on some or all of the keys implicit.
    \item Robustness depends on the ``entropy'' of the response. For instance, if the prompt asks for a completely deterministic response, then there is no way to embed a watermark in the first place --- let alone obtain robustness. We incorporate this dependence into our definition by only requiring the detector/decoder to function when a certain property $P$ of the model, prompt, and content is satisfied.
\end{itemize}
Because of the above issues, we present an extremely broad definition of robustness.
It is sometimes possible to prove that a scheme is robust to certain channels under certain conditions on the entropy of the response~\cite{zhao2024provable,kuditipudi2023robust,christ2024pseudorandom,golowich2024edit}, but watermark robustness is primarily evaluated empirically.

\begin{definition}[Robustness] \label{def:robustness}
    A watermark detector $\Detect$ is robust to a channel $\cE$ with error $\epsilon$ for property $P$ if, for any prompt $\prompt$,
    \ifthenelse{\boolean{@twocolumn}}{
        \begin{multline*}
            \Pr_{\substack{\gk,\dtk \\ x \leftarrow \Watermark_{\gk}^{\cM}(\prompt) \\ x' \leftarrow \cE(x)}}\Big[\Detect_{\dtk}(x') \to \false \text{ and } \\ P(\cM,\prompt,x) = \true \Big]
            \le \varepsilon.
        \end{multline*}
    }{
        \[
        \Pr_{\substack{\gk,\dtk \\ x \leftarrow \Watermark_{\gk}^{\cM}(\prompt) \\ x' \leftarrow \cE(x)}}\Big[\Detect_{\dtk}(x') \to \false \text{ and }  P(\cM,\prompt,x) = \true \Big]
                \le \varepsilon.
        \]
    }
    A watermark decoder $\Decode$ is robust to a channel $\cE$ with error $\epsilon$ for property $P$ if, for any message $\m$ and prompt $\prompt$,
    \ifthenelse{\boolean{@twocolumn}}{
        \begin{multline*}
            \Pr_{\substack{\gk,\dtk \\ x \leftarrow \Watermark_{\gk}^{\cM}(\m, \prompt) \\ x' \leftarrow \cE(x)}}\Big[\Decode_{\dck}(x') \not\to \m \text{ and } \\ P(\cM,\prompt,x) = \true \Big]
            \le \varepsilon.
        \end{multline*}
    }{
        \[
        \Pr_{\substack{\gk,\dtk \\ x \leftarrow \Watermark_{\gk}^{\cM}(\m, \prompt) \\ x' \leftarrow \cE(x)}}\Big[\Decode_{\dck}(x') \not\to \m \text{ and } P(\cM,\prompt,x) = \true \Big]
            \le \varepsilon.
        \]
    }
\end{definition}

The property $P$ typically measures some form of entropy of the content, ensuring that we do not require deterministic responses to be watermarked.
For instance, in \cite{christ2024undetectable}, $P(\cM, \pi, x)$ is $\true$ whenever the probability of sampling $x$ as a response to $\cM$ on prompt $\pi$ is sufficiently low.
Image watermarks also require entropy in the response, although this is usually not very limiting, and can be made robust to channels that introduce bounded error in the pixel/latent space.

While robustness is a desirable property of a watermark, even non-robust watermarks may be useful in many settings because it is likely that most users will not attempt to remove the watermark.
Such watermarks have a low \emph{false negative rate}, i.e., strong robustness against the identity channel $\cE(x) = x$.
For text, the false negative rate can often be computed precisely in terms of the ``entropy'' in the given text portion, where the notion of ``entropy'' depends on the scheme. 
The utility of non-robust watermarks is especially apparent for watermarking applications where the user does not suffer any negative consequences when caught: For example, one may wish to detect AI-generated images online in order to omit them from training datasets.

\subsection{Unforgeability} \label{subsec:unforgeability}
Thus far, we have only required that knowledge of the generation key \emph{allows} one to embed a reliable watermark.
In some settings, it is additionally desirable that knowledge of the generation key is \emph{required} in order to embed a watermark.
Such a guarantee, called \emph{unforgeability}, is important when using a watermark for \emph{attribution}: that is, determining with certainty that some content came from a model.
For instance, if one wishes to use a watermark to blame a model for producing harmful content, unforgeability is essential.

In more detail, one natural way to define unforgeability is to require that it be computationally infeasible for an attacker without knowledge of an embedding key to produce watermarked content that was not output verbatim by the watermarked model.
Note that, while unforgeability is related to the false positive rate, the latter says nothing about the success of a dedicated attacker intentionally trying to get their content flagged.

The first thing to observe is that this definition of unforgeability is fundamentally incompatible with robustness.
If a detector is robust, then it is possible to generate a response from the model, modify it slightly to make it problematic, and then falsely attribute the problematic content to the watermarked model, which could be undesirable in some cases.


However, this incompatibility can be circumvented by simply using separate algorithms for detection (which should be robust) and attribution (which should be unforgeable).
It is possible to construct a \emph{single} watermark generation algorithm such that later, at detection/attribution time, one can choose whether to use the detection or attribution algorithm depending on whether one desires robustness or unforgeability.
Therefore, we keep the attribution algorithm and key ($\Attribute$ and $\ak$) separate from the detection algorithm and key ($\Detect$ and $\dtk$). An alternative way would be to modify the unforgeability requirement so that it is computationally infeasible for an attacker without knowledge of an embedding key to produce watermarked content that is dissimilar --- rather than verbatim --- from the content generated by the watermarked model. Here, we focus on the first approach and refer the reader to~\cite{fairoze2023publicly} for more discussion on the second.


\begin{definition}[Unforgeability~\cite{fairoze2023publicly}]
A watermark is \emph{unforgeable} if for all $\lambda$ and polynomial-time algorithms $\cA$,
\ifthenelse{\boolean{@twocolumn}}{
    \begin{multline*}
        \Pr_{\substack{\gk, \ak \\ x \gets \cA^{\Watermark^\cM_\gk}(1^{\lambda}, \ak)}}\left[\Attribute_{\ak}(x) \to \true \text{ and } x \notin \cQ \right] 
        \\ \leq \mathsf{negl}(\lambda),
    \end{multline*}
}{\[
    \Pr_{\substack{\gk, \ak \\ x \gets \cA^{\Watermark^\cM_\gk}(1^{\lambda}, \ak)}}\left[\Attribute_{\ak}(x) \to \true \text{ and } x \notin \cQ \right] 
             \leq \mathsf{negl}(\lambda),
\]}
where $\cQ$ denotes the set of responses obtained by $\cA$ on its queries to the watermarked model.
\end{definition}
That is, $\cA$ queries the watermarked model and attempts to produce its own content $x$ that is watermarked but is distinct from any of the responses to its queries.

Using digital signatures, it is possible to make the attribution key $\ak$ public without sacrificing unforgeability.
In this case, we say that the scheme supports \emph{unforgeable public attribution}.
Such watermarking schemes are sometimes referred to as simply ``publicly detectable'' in the literature, but we avoid this terminology here as it does not specify which keys can be published, or which properties are preserved when they are.
Depending on the desired properties of the scheme, it may suffice to simply publish the watermarking key.
For instance, if the only desired properties are that the watermark preserves quality and has low false positive and negative rates, then there is no harm in publishing the key.

Conversely, there are other reasonable properties that could be desired of ``publicly detectable'' watermarks that are not implied by unforgeability.
For instance, we may wish to preserve robustness after publishing the detection key, or to preserve undetectability after publishing the generation key.

\subsection{Support for Embedding Messages}
If the watermark generation algorithm can embed messages in the content, then we say that it is a \emph{multi-bit watermark}.
Such watermarks are crucial for meeting the increasing demands of customization in various applications of generative models.
For instance, embedding detailed metadata, such as model version and user-specific information, enables efficient traceability of the text's origin across multiple models and users.
These capabilities are particularly vital in large-scale deployments.

In general, the length of the message that can be embedded in some given content depends on the amount of entropy in that content.
Intuitively, greater entropy gives the embedder more freedom in generating content that is consistent with its desired message.
Images, audio, and video tend to have high entropy, and therefore are more amenable to embedding nontrivial messages.
In contrast, text is relatively low entropy, and practical text watermarks typically do not support messages longer than a few bits.

\subsection{Computational Efficiency}
The watermarked generation process should impose minimal computational overhead, ensuring that the computational efficiency remains comparable to that of the unwatermarked generation process in terms of computation, memory usage, and throughput.
Significant computational complexity can hinder real-world deployment, making the model less appealing for practical applications.

The computational requirements of detection, decoding, and attribution should also be kept at a minimum.


\section{Threat Models}
\label{sec:threat_model}

%
A common concern with watermarking schemes is their robustness against various perturbations and attacks. This section defines the threat models associated with watermarking, focusing on attack objectives, as well as the adversary's knowledge and capabilities.
\subsection{Attack Objectives}
We identify two primary adversarial objectives in watermarking schemes: \emph{removal} and \emph{forgery}. Note that when pursuing these two objectives, the attacker should also strive to preserve the quality of the generated outputs. Otherwise, by compromising the quality, any watermarked (or non-watermarked) content could be trivially identified as non-watermarked (or watermarked).

\noindent\textbf{Watermark Removal.}
The watermark removal attack aims to modify AI-generated content such that the detection algorithm classifies it as \textit{not watermarked}, or the decoding algorithm extracts an incorrect message. This attack would effectively bypass the mechanisms designed to identify content originating from a specific model. The adversary removes watermarks embedded in generated content by introducing \emph{small} perturbations that \emph{distort} the watermarks while preserving content quality. Basic methods from prior studies include simple image manipulations, such as resizing or adding noise~\cite{an2024waves,zhao2024latent}, and text edits, such as random deletions or substitutions~\cite{kirchenbauer2023watermark, zhao2024provable}. However, more advanced attackers may develop specialized techniques to effectively remove watermarks while preserving the content's quality and coherence.

\noindent\textbf{Forging Watermarks.}
The objective of the forging attack is the opposite of watermark removal. The adversary seeks to create content that is falsely classified as \textit{watermarked}, despite not being processed by the designated watermarking scheme. This attack allows malicious actors to attribute their content to a specific model or system, potentially undermining trust and accountability mechanisms. Forgery can be achieved through various methods, from simpler approaches, such as collecting and analyzing watermarked content to reconstruct watermarks for future use, to more sophisticated attacks where knowledge of the target watermarking algorithm to embed them into any content of interest.

\noindent\textbf{Secret Extraction.}
A more advanced adversarial objective is the extraction of the secret keys used in the watermarking scheme. This attack is significantly harder than removal or forging attacks. In most cases, if an adversary successfully extracts the secret keys, they can easily remove or forge watermarks. However, it is important to note that extracting the secret keys is not a prerequisite for executing removal or forging attacks.

\subsection{Adversary Knowledge and Capabilities}

It is essential to explicitly define what systems and information an adversary is assumed to have access to. These capabilities determine the feasibility and success of various attack strategies. Here are several dimensions to consider:

\noindent\textbf{Generator Oracle Access.}
Does the adversary have access to generate additional outputs watermarked by the same algorithm with the same key? Access to such an oracle can help the adversary understand how watermarks are applied and create more effective removal or forging attacks.

\noindent\textbf{Access to Watermarked Content.}
If the adversary does not have oracle access to the watermarked generator, how many watermarked generations are available to them? Access to known watermarked contents can help in reverse engineering the watermarking process.

\noindent\textbf{Access to Non-Watermarked Content.}
If the adversary lacks oracle access to the watermarked generator, how many non-watermarked generations do they have access to? This capability helps the adversary understand the differences between watermarked and non-watermarked content.

\noindent\textbf{White-box Access to the Model.}
Does the adversary have knowledge of the generative model underlying the watermark scheme, and do they have white-box access to this model? Such access enables the adversary to modify internal parameters or the generation process itself. This capability can facilitate watermark removal or allow the model to be repurposed for alternative tasks.

\noindent\textbf{Non-Watermarked Generator Access.}
Does the adversary have access to query the model without any watermarks applied? This would enable them to compare outputs to detect the presence of a watermark.

\noindent\textbf{Chosen Key Oracle Access.}
Can the adversary generate watermarked outputs using chosen keys? This would help in forging watermarked outputs or learning the watermark embedding process.

\noindent\textbf{Verifier Feedback Granularity.}
What level of feedback does the verifier provide when queried by the adversary~(e.g., through $\Detect$, $\Decode$, or $\Attribute$)? Feedback can range from a simple true/false response to a detailed probability score. The granularity of the verification output can significantly influence the adversary’s ability to refine their strategy.

\noindent\textbf{Verifier Oracle Access.}
What level of access does the adversary have to the verifier? Can they request verification for specific outputs, and if so, how many times can they do this? Repeated access can provide the adversary with significant information on how the verifier works, and they can iteratively modify content until the verifier fails to remove the watermark.

\noindent\textbf{Surrogate Model Access.}
Does the adversary have access to another model to serve as a surrogate? Although such models may have inferior performance, these surrogate models can be employed as paraphrasing or editing tools to modify or regenerate content, potentially removing the watermark.

\section{Empirical Evaluation Methodologies} \label{sec:eval_method} 

This section outlines the standard practices for evaluating watermarks, with a particular focus on text and image watermarking. Key aspects of evaluation include detection effectiveness, robustness against attacks, and quality preservation.

\subsection{Detection Effectiveness}

The foremost property that a watermark should satisfy is that it is effectively detectable, while minimizing the risk of false-positives.
Typical metrics of interest include:

\begin{itemize}
    \item \textbf{AUROC (Area Under the Receiver Operating Characteristic Curve):} Measures the trade-off between true positive and false positive rates of watermark detection.
    \item \textbf{Fixed FPR Comparisons:} In many scenarios, a high false-positive rate is untenable (e.g., when using watermarking to detect misinformation). It is thus standard to also evaluate the detection performance for a fixed, low false positive rate (e.g., at 0.1\%).
\end{itemize}

The detection performance can often be computed analytically. Many watermarking works derive explicit theoretical bounds on their false-positive and false-negative rates.
However, it is also good practice to complement these formal bounds with empirical measurements, for two reasons. First, bounds on the false negative rate (i.e., the likelihood that watermarked content is undetected) often depend on statistical characteristics of the content, such as the entropy of a piece of text. Second, inherent randomness in machine learning models and system components (e.g., GPUs) can slightly reduce the detectability of a watermark.

\subsection{Robustness Against Attacks}

Robustness is a critical metric for evaluating watermark effectiveness, particularly in the face of adversarial manipulations. Below, we categorize different types of attacks commonly used to evaluate robustness.


\subsubsection{Evasion Attacks}
The primary threat faced by watermarks and which should always be evaluated rigorously is an evasion attack, where an adversary removes the watermark from a piece of AI generated content.
We distinguish between three coarse categories of attacks that watermarking schemes are typically evaluated against:

\begin{itemize}
    \item \textbf{``Edit Attacks'':} these attacks make ``small'', typically localized changes to a watermarked content, in the hope of removing the watermark. It is a form of ``noisy'' channel in the sense of Definition~\ref{def:robustness}.
    
    Common examples include: (1) \emph{text deletion attacks} that remove a fraction of words from a generated text; (2) \emph{image distortion attacks} that apply simple transformations or noising to generated images or videos; (3) \emph{text replacement attacks} that change individual words with synonyms.

    Depending on the threat model, these attacks can make use of automated optimization procedures. For example, an attacker with white-box access to a watermark detector could optimize adversarial noise that fools the detector~\cite{jiang2023evading}.
    An attacker could also first ``steal'' or ``estimate'' the watermarking pattern~\cite{jovanovic2024watermark, yang2024steganalysisdigitalwatermarkingdefense} and then make targeted edits to remove the watermark.

    \item \textbf{``Regeneration Attacks'':} these attacks pass a watermarked output through a different (non-watermarked) generative AI model, in the hope of removing the watermark while preserving the output's utility.
    This attack can also be viewed as a ``noisy'' channel in Definition~\ref{def:robustness}.

    The most common examples include applying a paraphrasing or summarization model to a piece of watermarked text~\cite{zhang2023watermarks, piet2023mark, pan2024markllm}, or applying a denoising autoencoder or diffusion model to a watermarked image~\cite{zhao2024latent, saberi2024latent}.
    
    \item \textbf{``Downsampling Attacks'':} these attacks create a watermarked output that contains the actual desired output as a subset. Extracting this subset can then destroy the watermark. 
    This attack differs from the above attacks in that it explicitly \emph{deletes} rather than replaces part of the watermarked output. It can thus be viewed as a ``deletion'' channel in the sense of Definition~\ref{def:robustness}.

    A notable example is the \textit{Emoji Attack} (or \textit{Pineapple Attack}), where the attacker prompts a language model to embed special symbols (e.g., emojis) between words. Removing these special words after generation significantly disrupts watermarks, especially those based on $k$-gram statistics.
    
    A similar approach could be considered for image watermarks, although we are unaware of such an attack being evaluated in prior work. For example, the attacker could ask for a generation of ``an image with object X on the left and object Y on the right'', and then simply crop half of the output. Depending on the watermarking scheme, this may significantly weaken the watermark.

\end{itemize}

\subsubsection{Forging Attacks}
Forging or spoofing attacks aim to create content that can be falsely claimed as watermarked.  In principle, edit attacks based on adversarial optimization can be ``reversed'' to yield a forgery attack (unless forgery is computationally infeasible). Notably, most forging strategies do not require knowledge of the exact watermark key. For instance,~\cite{gu2023learnability, jovanovic2024watermark} demonstrate that with access to a substantial corpus of both watermarked and unwatermarked text, an adversary can (partially) learn the red-green watermarking rule of Kirchenbauer et al.~\cite{kirchenbauer2023watermark}, and then produces text that either maximizes (for forgery) or minimizes (for evasion) the approximate watermarking score. Similarly, Saberi et al.~\cite{saberi2024latent} demonstrate that adversaries can train a surrogate model using watermarked and non-watermarked images, then apply adversarial attacks~(e.g., PGD~\cite{madry2017towards}) to make non-watermarked images appear watermarked. And such attacks often transfer effectively to the true watermark detector.

\subsection{Quality Assessment}

Watermarking techniques must preserve the quality of the original content, whether in text or image formats.

\subsubsection{Text Quality}
The quality of watermarked text is typically evaluated by comparing it to unwatermarked text generated by the same language model. Perplexity (PPL) is a widely used metric for measuring the fluency of generated text by assessing its probability under a language model. Diversity is another critical metric, which can be evaluated through \(n\)-gram repetition rates, Distinct \(n\)-grams, and Self-BLEU~\cite{zhu2018texygen}. Self-BLEU calculates the BLEU score for each generated sentence by considering other generated sentences as references, where lower Self-BLEU scores indicate higher diversity and less repetition.

MAUVE Score~\cite{pillutla2021mauve} is often employed to quantify the distributional similarity between watermarked model outputs and human-written text. Additionally, LLM-as-a-Judge~\cite{zheng2023judging} can be leveraged to compare the quality of watermarked and unwatermarked content, determining human-like preferences through win-rate analysis. Human evaluation, involving subjective scoring or ranking by human assessors, is another critical approach to assess the clarity, coherence, and overall quality of watermarked text.

For task-specific evaluations, metrics such as ROUGE~\cite{lin2004rouge}, BLEU~\cite{papineni2002bleu}, BERTScore~\cite{zhang2019bertscore}, BARTScore~\cite{yuan2021bartscore}, and InstructScore~\cite{xu2023instructscore} are used to measure the semantic and lexical alignment for summarization and translation tasks. In the context of code generation, executable accuracy metrics like Pass@k are used to determine the correctness of generated code.

\subsubsection{Image Quality}
The quality evaluation of watermarked images depends on whether the watermark is applied during or after the image generation process. Metrics such as Peak Signal-to-Noise Ratio (PSNR) and Structural Similarity Index (SSIM) are commonly used to measure quality degradation in post-generation watermarking. For assessing fidelity and perceptual similarity in generative models, metrics like Frechet Inception Distance (FID)~\cite{heusel2017gans}, CLIP Score~\cite{radford2021learning}, Inception Score~\cite{salimans2016improved}, and LPIPS~\cite{zhang2018unreasonable} Score are employed.

Human evaluation also plays a significant role in image quality assessment. Reviewers are asked to rank or score watermarked images based on perceived quality, clarity, and naturalness. In addition to these methods, advanced metrics such as DreamSim~\cite{fu2023dreamsim}, which evaluates fine-grained perceptual similarity, BLIP Score~\cite{li2022blip}, which combines vision and language models to assess image-text alignment, and ImageReward~\cite{xu2024imagereward}, which measures subjective preferences based on human-labeled datasets, have emerged as valuable tools in image quality assessment.




\section{Recent Representative Watermark Schemes} \label{sec:represent_works} 
In this section, we present several representative works in the field to illustrate how watermarking schemes operate. We discuss the strengths and limitations of different methods to provide a balanced understanding of their effectiveness. We also connect these representative works with the desired properties of a watermarking scheme discussed in Section \ref{sec:watermark_method}. An approximate conceptual comparison is shown in Table \ref{tab:wm-comparison}.


\subsection{Text Watermarks}
Early text watermarking approaches primarily applied post-processing techniques (add watermarks to existing text), which can be categorized into format-based, lexical-based, and syntactic-based methods~\cite{liu2024survey}.
Format-Based watermarking alters text presentation, such as line shifts, Unicode adjustments, or whitespace variations (e.g., UniSpaCh~\cite{por2012unispach}, EasyMark~\cite{sato2023embarrassingly}). These methods, while simple, are susceptible to removal through reformatting or canonicalization. Lexical-based watermarking, on the other hand, replaces specific words with synonyms while retaining sentence structure. This technique, which employs resources like WordNet~\cite{fellbaum1998wordnet} or context-aware models (e.g., BERT-based infill~\cite{devlin2018bert}), enhances robustness by preserving semantic integrity against attacks. Syntactic-based watermarking modifies sentence structure through techniques like adjunct movement, clefting, or passivization~\cite{atallah2001natural}. Although language-dependent, syntactic-based watermarking is generally more resilient to basic attacks due to its structural approach.

In this paper, we focus more on advanced watermarking approaches that integrate watermarking directly into text generation process. Below, we discuss recent representative works in this domain.


\noindent\textbf{Green-Red Watermark.}
Green-Red Watermark embeds a watermark message into logits generated by LLMs without altering model parameters. It modifies logits by partitioning the vocabulary at each token position into a red list (\( R \)) and a green list (\( G \)). These partitions may be specified with the secret key and fixed for all tokens as in~\cite{zhao2024provable}, or chosen pseudorandomly as the response is generated as in~\cite{kirchenbauer2023watermark}. Specifically,~\cite{kirchenbauer2023watermark, kirchenbauer2024reliability} choose these partitions based on a pseudorandom function (i.e., PRF or hash function) applied to the preceding $k$-gram.
For each token \( i \) generated by the watermarked model, a bias \( \delta \) is applied to the logits of tokens in \( G \), yielding adjusted logits \( \tilde{l}(i)_j \), which are computed as \( \tilde{l}(i)_j = l(i)_j + \delta \) if \( v_j \in G \), and \( \tilde{l}(i)_j = l(i)_j \) if \( v_j \in R \). 
This adjustment biases the generation toward green tokens, increasing their proportion in the watermarked text. 
Detection of the watermark involves categorizing tokens as red or green using the same hash function and calculating the green token ratio with the \( z \)-metric, defined as \( z = \frac{|s|_G - \gamma T}{\sqrt{T \gamma (1 - \gamma)}} \), where \( T \) is the text length, \( |s|_G \) is the number of green tokens, and \( \gamma \) is the predefined green token ratio. 
A text exceeding a certain \( z \)-metric threshold is classified as watermarked. The biasing strategies of the above Green-Red watermark result in nonzero distortion for each token.

Numerous studies have explored variations and improvements to Green-Red Watermarking, including \cite{lee2023wrote, liu2023unforgeable,liu2024adaptive,huo2024token, zhou2024bileve, qu2024provably}, among others.

\newcommand{\levelcircle}[1]{%
  \tikz[baseline=-0.5ex] \filldraw[fill=black!#1, draw=black] (0,0) circle (0.12cm);
}

\begin{table}[t]
\centering
\small
\resizebox{\linewidth}{!}{
\begin{tabular}{lcccccc}
\toprule
\textbf{Method} & \textbf{Quality} & \textbf{FPR} & \textbf{Robustness} & \textbf{Unforgeability} & \textbf{Messages} & \textbf{Efficiency} \\
\midrule
\multicolumn{7}{l}{\textit{Text Watermarking}} \\
Green-Red WM \cite{kirchenbauer2023watermark,zhao2024provable} & \levelcircle{50} & \levelcircle{100} & \levelcircle{100} & \levelcircle{25} & \levelcircle{0} & \levelcircle{100} \\
Gumbel WM \cite{aaronson} & \levelcircle{75} & \levelcircle{100} & \levelcircle{50} & \levelcircle{75} & \levelcircle{0} & \levelcircle{100} \\
Undetectable WM \cite{christ2024undetectable} & \levelcircle{100} & \levelcircle{75} & \levelcircle{25} & \levelcircle{100} & \levelcircle{0} & \levelcircle{75} \\
Pseudorandom Codes \cite{christ2024pseudorandom} & \levelcircle{100} & \levelcircle{75} & \levelcircle{50} & \levelcircle{100} & \levelcircle{25} & \levelcircle{50} \\
Semantic Sentence WM \cite{hou2023semstamp} & \levelcircle{75} & \levelcircle{50} & \levelcircle{75} & \levelcircle{50} & \levelcircle{0} & \levelcircle{25} \\
\midrule
\multicolumn{7}{l}{\textit{Image Watermarking}} \\
Stable Signature \cite{fernandez2023stable} & \levelcircle{50} & \levelcircle{100} & \levelcircle{25} & \levelcircle{25} & \levelcircle{100} & \levelcircle{100} \\
Tree-Ring WM \cite{wen2023tree} & \levelcircle{50} & \levelcircle{100} & \levelcircle{75} & \levelcircle{50} & \levelcircle{0} & \levelcircle{100} \\
Gaussian Shading WM \cite{yang2024gaussian} & \levelcircle{75} & \levelcircle{100} & \levelcircle{100} & \levelcircle{25} & \levelcircle{100} & \levelcircle{100} \\
PRC Watermark \cite{gunn2024undetectable} & \levelcircle{100} & \levelcircle{100} & \levelcircle{100} & \levelcircle{100} & \levelcircle{100} & \levelcircle{100} \\
\bottomrule
\end{tabular}
}
\caption{Qualitative comparison of representative watermarking methods across key evaluation criteria. Darker circles indicate stronger performance, offering an intuitive view of trade-offs without numerical scores.} 
\label{tab:wm-comparison}
\end{table}

\noindent\textbf{Gumbel Watermark.} Aaronson~\cite{aaronson} proposed a watermark that, like Kirchenbauer et al.~\cite{kirchenbauer2023watermark}, biases its choice of the next token based randomness derived from a PRF applied to the preceding $k$-gram.
Notably, the \emph{Gumbel} biasing strategy used by Aaronson~\cite{aaronson} results in a computationally \emph{distortion-free} watermark, provided that $k$ is large enough that no $k$-gram is repeated.
That is, the distribution of a single watermarked response is indistinguishable from that of a single response from the original model.

Kuditipudi et al.~\cite{kuditipudi2023robust} also apply the Gumbel sampler to construct a distortion-free watermark, but use fixed randomness to yield stronger robustness. 
That is, instead of applying a PRF to $k$-grams, they cycle through a fixed, pre-determined sequence of values called the ``key.'' 
To detect watermarked text, 
one iterates over all possible key sequences and test for similarity between the key and the given response.
If there are $L$ key sequences, this brute-force search increases the computational complexity of the detector and false positive rate by a factor of $L$.
However, this detection strategy also allows them to tolerate edits up to a constant rate of changes.

Variants of Gumbel Watermark techniques have been explored further in~\cite{hu2023unbiased,fernandez2022watermarking,wu2024resilient,zhao2024permute,fu-etal-2024-gumbelsoft}, among others.

\noindent\textbf{Undetectable Watermark.} 
Christ et al.~\cite{christ2024undetectable} construct an undetectable watermark.
Similarly to Aaronson~\cite{aaronson}, they use a PRF applied to preceding tokens in order to derive randomness used to bias the next token in a distribution-preserving way.
Their key idea in achieving undetectability is ensuring that the token sequences input to the PRF are \emph{never repeated}.
They do so by only using a sequence as a PRF input once it contains enough \emph{empirical entropy}.
Empirical entropy quantifies how much randomness is in the given sequence, and it can be computed given the relevant logits.

In more detail, this undetectable watermark outputs tokens $v_1, \ldots, v_\ell$ sampled from the original model until it has observed enough empirical entropy, relative to some fixed threshold.
At this point, it uses $(v_1, \ldots, v_\ell)$ as an input to a PRF, and uses the resulting randomness to bias the remaining tokens in the response.
Choosing $v_1, \ldots, v_\ell$ to contain enough empirical entropy ensures that PRF inputs are never repeated, and therefore the watermark introduces no correlations across tokens.
However, if any token of $(v_1, \ldots, v_\ell)$ is changed, the watermark is removed.

One can make this watermark robust to cropping, by applying it in ``blocks'': selecting a new PRF input once a detectable signal has been embedded using the previous PRF input.
However, even this cropping-robust watermark is not at all robust to token substitutions.

In follow-up work, Fairoze et al.~\cite{fairoze2023publicly} constructed an undetectable watermark with unforgeable public attribution.
The watermark relies on embedding digital signatures in LLM-produced text such that it is computationally impossible to forge watermarked text without including a high proportion of contiguous tokens from honestly-watermarked text.
Their scheme can easily be extended to one that satisfies the definition of unforgeable public attribution presented here, while still allowing for a separate detection algorithm that is robust.

\noindent\textbf{Pseudorandom Error-correcting Codes.} Christ \& Gunn \cite{christ2024pseudorandom} use \emph{pseudorandom error-correcting codes (PRCs)} to construct the first language model watermark that is both undetectable and robust to a constant fraction of token substitutions.
Codewords of a pseudorandom error-correcting code both appear random and are robust.
This watermarking scheme samples each LLM response so that its tokens have significant correlation with a codeword.
To detect the watermark, one applies the PRC decoder to the response.
Robustness of the codeword ensures that as long as the response has not been modified too much, the decoder succeeds and the watermark is detected.
Pseudorandomness of the codeword ensures that sampling the response to be correlated with this codeword introduces no noticeable bias, which implies undetectability. Further advancements in PRCs are discussed in~\cite{cohen2024watermarking,golowich2024edit,ghentiyala2024new}, although PRCs have yet to be implemented for a practical LLM watermarking scheme.

\noindent\textbf{Semantic Sentence Watermark.} Several studies have explored sentence-level semantic watermarks, which enhance robustness against token-level modifications by treating sentences as the fundamental unit. SemStamp~\cite{hou2023semstamp} uses locality-sensitive hashing (LSH)~\cite{indyk1998approximate} to map candidate sentences into a semantic watermark space and employs rejection sampling to ensure generated sentences fall within a valid semantic region. Detection involves a one-proportion $z$-test on the number of valid-region sentences, with robustness enhanced by a margin-based constraint. However, its sequential sampling process is inefficient and fails to utilize prior semantic information. k-SemStamp~\cite{hou2024k} improves robustness by replacing LSH with k-means clustering, incorporating clustering results as intrinsic semantic information. Despite lacking theoretical guarantees, these methods offer insights for advancing semantic text watermarking.

\subsection{Image Watermarks}
Analogous to text watermarking, image watermarking techniques are divided into post-processing and in-processing schemes. Post-processing methods, traditionally favored for their wide applicability, embed watermarks into images after they have been generated. These techniques often operate in the frequency domain, modifying the image through transformations like Discrete Wavelet Transform~(DWT), Discrete Cosine Transform (DCT)~\cite{al2007combined}, and DWT-DCT-SVD~\cite{navas2008dwt}. Additionally, deep learning-based encoder-decoder approaches, including HiDDeN~\cite{Zhu2018HiDDeNHD}, RivaGAN~\cite{zhang2019robust}, StegaStamp~\cite{tancik2020stegastamp}, and SSL Watermarking~\cite{fernandez2022watermarking}, employ neural networks to embed and decode watermarks effectively.  In contrast, in-processing methods directly influence the generative model or the sampling process itself, embedding watermarks inherently within the generated images. 

This paper emphasizes the more advanced, in-processing watermarking approaches for image generation. Recent representative works are discussed below.

\noindent\textbf{Stable Signature.} Stable Signature~\cite{fernandez2023stable} embeds watermarking into the generation process by fine-tuning model parameters without altering the architecture. It modifies a pre-trained Latent Diffusion Model (LDM)~\cite{rombach2022high} so that all generated images encode a specific binary signature. A pre-trained watermark extractor $\cW$ retrieves the signature, and a statistical test verifies its origin. The method involves two steps: (1) pre-training $\cW$ to recover binary messages and (2) fine-tuning the LDM decoder $\cD$ to ensure all outputs embed a fixed signature $\m$. Similar approaches, like DiffusionDM~\cite{zhao2023recipe}, also fine-tune diffusion models to embed watermarks. However, these methods degrade generation quality, lack robustness to regeneration attacks, and offer no theoretical guarantees~\cite{an2024waves}.

\noindent\textbf{Tree-Ring Watermark.}
The Tree-Ring watermark, introduced by Wen et al.~\cite{wen2023tree}, modifies the latent sampling distribution of LDM and uses an inverse diffusion process for detection. The method fixes concentric rings in the Fourier domain of the latent space to zero. Detection involves applying DDIM inversion~\cite{songdenoising} to estimate the initial latent, and the watermark is considered present if the latent estimate exhibits unusually small values in the watermarked rings.
Subsequent works have refined this heuristic latent pattern~\cite{zhang2024robust, ci2024ringid}. However, the Tree-Ring watermark introduces significant deviations from the Gaussian latent distribution, resulting in reduced image quality and variability. Furthermore, it is a zero-bit scheme, incapable of encoding messages. While robust against certain attacks, it is vulnerable to adversarial surrogate attacks, as the latent pattern is easily learnable by neural networks~\cite{saberi2024latent}.

\noindent\textbf{Gaussian Shading Watermark.}
The Gaussian Shading watermark~\cite{yang2024gaussian} embeds a watermark by restricting latent space sampling to a fixed quadrant defined by the watermarking key. During detection, the latent is recovered, and its proximity to the watermarked quadrant is evaluated. Yang et al.~\cite{yang2024gaussian} claim ``lossless performance,'' supported by a proof that the distribution of a single watermarked image matches that of an un-watermarked image. However, this proof does not account for correlations across multiple generated images, which are critical for metrics such as FID~\cite{heusel2017gans}, CLIP Score~\cite{radford2021learning}, and Inception Score~\cite{salimans2016improved}.
To align with practical use cases, the same random watermarking key is used to generate multiple images for quality evaluation. As all images originate from the same quadrant in latent space, this approach inherently reduces variability, impacting diversity metrics.

\noindent\textbf{PRC Watermark.}
The PRC watermark~\cite{gunn2024undetectable} is an undetectable watermarking scheme designed for latent diffusion models. It employs a similar method to the Gaussian Shading watermark, except that a pseudorandom error-correcting code (PRC) is used to sample a fresh quadrant of latent space for each generation.
The main advantage of this approach is that, as a corollary of undetectability, it does not degrade the image quality --- even when measured across generations as in the FID, CLIP, or Inception Score.
Robustness of the pseudorandom code translates to robustness of the watermark.
This approach also enables messages to be embedded in the watermark.

\subsection{Video and Audio Watermarks}

Compared to the advancements in text and image watermarking, relatively few studies focus on video and audio watermarks specifically designed for generative AI. For video watermarking, existing methods often rely on traditional video watermarking techniques~\cite{li2001Overview, asikuzzaman2017overview} or apply post-processing watermarks to individual frames~\cite{zhang2019robust, luo2023dvmark}, similar to image watermarking. In the case of audio watermarking, many approaches embed watermarks into synthetic audio using a trained/fine-tuned watermark encoder, primarily targeting text-to-speech generative models~\cite{cho2022attributable, chen2023wavmark, tong2024enhancing, liu2023detecting, san2024proactive, liu2024groot, liu2024audiomarkbench}. Looking forward, watermarking for generative media could benefit from designs tailored to any-to-any generation pipelines~\cite{team2024chameleon}, including cross-modal generation~\cite{zhang2024v2a} and even other modalities such as tabular data generation~\cite{borisov2022language}.


\section{Open Problems and Discussion} \label{sec:open_prob}


As the field of watermarking for GenAI continues to evolve, there are several exciting opportunities for future research and development. This section delves into some of these open questions and discussions.

\noindent\textbf{Robustness and Unforgeable Public Attribution.} Who should have the right to detect watermarked content? In the most democratized case, the public attribution setting, watermarks can function analogously to digital signatures where a private entity is allowed to embed watermarks and anyone can detect their presence. From a technical perspective, it is possible to construct watermarks with unforgeable public attribution: the open accessibility of the detector does not introduce forgery attacks. 
However, present constructions either lack strong robustness or concrete efficiency. It remains open whether schemes with public attribution and strong robustness can be efficiently instantiated or if more powerful robustness notions can coexist with public attribution in theory.

\noindent\textbf{Trade-offs Among Different Properties.} In Section \ref{sec:watermark_method}, we discussed the desired properties of watermarks across six dimensions. Optimal watermarking schemes should aim to achieve a Pareto-optimal balance among these dimensions, both empirically and theoretically, while relying on minimal assumptions. Existing studies~\cite{huang2023towards, li2024statistical} claim optimality within certain restricted settings. We hope to see future research further expand these boundaries and push the limits of optimality.


\noindent\textbf{Watermarks for Copyright.}
It is tempting to use watermarks to \emph{prove} that a certain model generated some content, to claim copyright.
For example, if a user falsely claims they drew an AI-generated image themselves, the model owner may wish to reveal their secret watermark key to ``prove'' that their model in fact generated it.
However, for many existing schemes, a malicious model owner could take a user-generated image, and manufacture a \emph{fake} watermark key under which the image is watermarked!
Therefore, revealing this key is not convincing proof.

Existing formally defined watermark properties do not address this scenario.
A low false negative rate says simply that content that is produced independently of the secret watermarking key is unlikely to be watermarked.
However, we need a stronger property: It should be infeasible for an adversary, given some content, to produce a watermark key under which that content is detected.
For certain existing schemes, one can satisfy this definition by forcing the model owner to publish a commitment to their watermark key. 

A fruitful direction for future work is formalizing this setting of proof-of-generation, and showing how to achieve the desired guarantees.

\noindent\textbf{Semantic Watermark.}
For image generation, latent space watermarks are essentially semantic watermarks~\cite{wen2023tree, gunn2024undetectable}. As long as the latent embeddings remain relatively unchanged, the watermark persists. While semantic sentence watermarking in LLMs has seen recent progress (Section \ref{sec:represent_works}), paragraph or document-level semantic watermarking remains an open challenge. Ideally, watermarking the outline or structure of text (e.g., story ideas) could offer defense against strong attacks (e.g., translation attacks), making it a promising future direction.

\noindent\textbf{Is It Possible to Make Open-Source Models Watermarkable?} Existing watermarking methods often rely on modifying the generation process of LLMs or diffusion models. However, the open-source nature of many models makes this approach challenging, as users can employ various decoding methods. A more promising approach involves training models to inherently embed watermarks into their parameters. This would allow for watermark detection without modifying the generation process. While this approach presents significant technical challenges, it holds the potential for robust and scalable watermarking in open-source models.


\noindent\textbf{Other Applications of Watermarking.} Watermarking extends beyond content traceability and offers applications in various aspects of AI model and data governance. For instance:

\begin{itemize}
    \item \textbf{Model Watermarking:} Embedding a watermark within a model can help protect intellectual property by preventing unauthorized fine-tuning, distillation, or quantization of the model. This ensures that any modified version can still be traced back to the original owner, safeguarding ownership claims~\cite{he2022cater, zhao2022distillation, zhao2023protecting}.
    \item \textbf{Dataset Watermarking:} Datasets used for model training can also be watermarked to establish ownership and detect improper usage. For example, if a dataset is used for training without proper authorization or payment, a watermark can provide evidence of misuse, enabling legal or contractual enforcement~\cite{guo2024domain, wei2024proving}.
\end{itemize}

These applications highlight the broader potential of watermarking as a tool for ensuring ethical and accountable AI practices.

\noindent\textbf{Potential Misuse and Privacy Violations.}
The ways in which watermarks can be used maliciously have been underexplored.
Watermarks can be embedded in user-generated content \emph{without the user's knowledge or consent}, and enabling significant privacy violations.
For example, suppose a model provider embeds a user's ID in images that that user generates---practical image watermarking schemes are capable of embedding this kind of information.
If that user later posts that image on an anonymous but public social media account, anyone with detector access can link this account to the user's identity.
Traditional watermarks and fingerprints have also been used to track whistleblowers who leak emails---each recipient receives an imperceptibly different version of the email, e.g. with different whitespace.
One can use these imperceptible patterns to identify the source in the event of a leak.
Such tracking only becomes easier, and provably imperceptible, with undetectable and distortion-free watermarks.
Therefore, one should consider how generative AI watermarks introduce new privacy threats as well as intensify existing ones.

\noindent\textbf{Efficient Key Replacement.}
In many scenarios it may be desirable for the watermarking scheme to support efficient replacement of the watermarking keys.
For instance, if the generation key is compromised for a scheme that supports unforgeable attribution, then the generation and attribution keys will need to be replaced in order to restore unforgeability.
Most existing schemes support efficient key replacement, because one can efficiently sample fresh keys and simply begin using them.
However, this property may not be satisfied by schemes that involve re-training or fine-tuning a model to accommodate the watermark.

\noindent\textbf{Policy Challenges and the Future of Watermarking.} For watermarking to be universally effective in identifying AI-generated content, it requires widespread adoption by all organizations offering generative AI services. A unified watermarking scheme would enable comprehensive detection of AI-generated content across platforms. Without such standardization, detection efforts are likely to be fragmented, with model owners only capable of identifying outputs generated by their own systems. Achieving this level of coordination poses significant challenges. A number of measures may be necessary to establish and enforce universal watermarking standards – including open collaboration among companies coupled with governmental legislation and global policy frameworks. Such measures would ensure that AI-generated content remains traceable, thereby addressing concerns related to misinformation, intellectual property infringement, and ethical accountability. The future of watermarking depends not only on technological advancements but also on cohesive policy efforts across international boundaries. It is important that these policy efforts are aligned with state-of-the-art watermarking schemes, cognizant of their differences, capabilities, and limitations~\cite{christodorescu2024securing}.
\section{Conclusion}

In summary, watermarking for AI-generated content offers a promising solution to the challenge of distinguishing between human and AI-created outputs. This paper has provided a comprehensive overview of the field, covering motivations, definitions, properties, threat models, evaluation strategies, and recent advancements. Future research could focus on novel algorithms leveraging advanced cryptographic, information-theoretic, and statistical principles to improve watermark performance.  Collaboration among researchers, policymakers, and industry stakeholders remains essential to address ethical implications and ensure the responsible deployment of watermarking in the evolving generative AI landscape.

\newpage
\bibliography{ref_paper}
\bibliographystyle{unsrt}


\end{document}